
\magnification=\magstep1
\hoffset=1.92 truecm
\voffset=1.2  truecm
\hsize=15.2 truecm
\vsize=21.7 truecm
\parindent=1.0 truecm
\parskip=0pt
\nopagenumbers
\def\cambaselines{\baselineskip=10.10pt
                  \lineskip=0pt
                  \lineskiplimit=0pt}
\def\oneskip{\vskip\baselineskip}
\cambaselines
\def\endmode{\par\endgroup}
\def\title{\begingroup \raggedright \noindent  \bf }
\def\endtitle{\endmode \oneskip\oneskip\oneskip\oneskip}
\def\authors{\begingroup \parindent=2.0truecm \obeylines }
\def\endauthors{\endmode \oneskip\oneskip}
\outer\def\beginsection#1\par{\vskip0pt plus.1\vsize\penalty-250
\vskip0pt plus-.1\vsize \bigskip\vskip\parskip
\message{#1}\leftline{#1}\nobreak\oneskip\noindent}
\centerline{\null}
\oneskip \oneskip \oneskip
\def\prb{Phys. Rev. B}
\def\prl{Phys. Rev. Lett.}

\def\half{{1\over 2}}
\def\cA{{\bf \cal A}}
\def\cL{{\bf \cal L}}
\def\bA{{\bf A}}
\def\bn{{\bf n}}
\def\cD{{\bf \cal D}}
\def\cO{{\bf \cal O}}
\def\cF{{\bf \cal F}}
\def\cH{{\bf \cal H}}
\def\bJ{{\bar J}}
\def\bt{{\bar t}}
\def\bS{{\bf S}}

\def\bk{{\bf k}}

\def\ijc{{\langle i ; j \rangle}}
\def\ijkc{{\langle i ; jk \rangle}}

\def\CuO2{{\rm CuO_2}}

\def\y1{{\rm YBa_2 Cu_2 O_7}}
\def\y6{{\rm YBa_2 Cu_2 O_6}}
\def\Tc{{\rm T}_c}
\def\OM{{\hat \Omega}}


\def\cA{{\bf \cal A}}

\def\cD{{\bf \cal D}}
\def\cF{{\bf \cal F}}
\def\cH{{\bf \cal H}}
\def\bS{{\bf S}}

\def\bk{{\bf k}}
\def\bfx{{\bf x}}

\def\ijc{{\langle i ; j \rangle}}
\def\ijkc{{\langle i ; jk \rangle}}



\title

THE SEMICLASSICAL EXPANSION OF THE T-J MODEL

\endtitle


\authors

Assa Auerbach
Department of Physics
Technion-IIT, Haifa 32000, Israel.

\endauthors


\beginsection 1. INTRODUCTION

The discovery of high temperature
superconductivity has spurred intense investigations of the two dimensional
doped
antiferromagnet.
In the strong coupling limit, the t-J Hamiltonian, derived from the large-U
Hubbard model, is often used to describe  the low lying excitations. At
zero doping, it directly reduces to the quantum antiferromagnetic Heisenberg
model
(QHM). Substantial progress has been recently
achieved in understanding the Heisenberg limit, both theoretically and
experimentally [C1].  The effects of doping, however, are still highly
controversial. As demonstrated in the other lectures of this Winter School, the
problem of interacting spins and charges requires novel theoretical approaches.

Let us start from the simplest one-band form of the Hubbard Model:
$$
\cH ~= -\sum_{\ijc,s} t  c^\dagger_{is}
c_{js} ~+
U\sum_{i}n_{i \uparrow} n_{i \downarrow}
\eqno(1.1)
$$
where $c^\dagger_{is}$ creates an electron at site $i$ with spin index $s$, and
$~\ijc$ denotes a summation over all sites and their
nearest neighbors on the square lattice. The large-U limit of Eq. (1.1) can be
derived using second order perturbation theory in $t/U$. In the restricted
Hilbert space of no double occupancies, the lowest order effective Hamiltonian
is the t-J model:
$$
\eqalign{
\cH^{tJ}~&=P_s \Bigg[ -\sum_{\ijc,s} t  c^\dagger_{is}
c_{js}\cr
&~~~- {J \over 4} \sum_{\ijkc}  (c^\dagger_{i\uparrow}
c^\dagger_{j\downarrow}-c^\dagger_{i\downarrow} c^\dagger_{j\uparrow})
(c_{j\downarrow} c_{k\uparrow}-c_{j\uparrow} c_{k\downarrow})
\Bigg] P_s~+\cO(t^3/ U^2)\cr}
\eqno(1.2)
$$
where $J=4t^2/U$ is the ``{\it superexchange}'' constant, and  $P_s$ projects
onto the states with no double occupancies. $\ijkc$ are triads of nearest
neighbors.
At half filling, (one electron per site), Eq. (2) reduces to  the
antiferromagnetic Quantum Heisenberg Model (QHM) on the square lattice:
$$
\cH^{tJ}\Bigg|_{n_i=1}~\to \cH^{QHM}~=
 {J\over 2} \sum_{\ijc} \left(\bS_i \cdot \bS_j - {1 \over 4}
\right)
\eqno(1.3)
$$
where
$$
\bS^\alpha_i~\equiv \half \sum_{ss'} c^\dagger_{is} \sigma^\alpha_{ss'}
c^\dagger_{is'}~~~~,~~~\alpha=1,2,3
\eqno(1.4)
$$
are spin 1/2 operators in the subspace of one electron per site, and
$\sigma^\alpha$ are the three Pauli matrices.
Charge fluctuations are completely suppressed at half filling, and we are left
with the spin excitations of the Heisenberg model. The two dimensional quantum
Heisenberg model has been successfully addressed by two complementary methods:
the semiclassical approximation (large-S expansion) [C1], and the Schwinger
boson mean field theory (large-N expansion) [A1]. There are also Fermion
large-N expansions which do not recover the known properties
of the QHM on the square lattice, and we shall not review them here.
In order to address the doped system in a formulation than readily recovers
the Heisenberg limit, we introduce two commuting Schwinger bosons $a_i,b_i$,
and a slave fermion $f_i$, to
represent the allowed states of the projected Hilbert Space. The operators obey
$$
\eqalign{
[a_i,a^\dagger_j]~=\delta_{ij} ~~~~,&~~~~\{f_i,f^\dagger_j\}~=\delta_{ij}~,\cr
[a_i,b_j^{(\dagger)}]~=0  ~~~~,&~~~~{\rm etc.}\cr
}
\eqno(1.5)
$$
and satisfy the local constraints
$$
a_i^\dagger a_i+ b_i^\dagger b_i+ f_i^\dagger f_i~=1~~~~~\forall ~i.
\eqno(1.6)
$$
The t-J model is faithfully represented by
$$
\cH^{tJ} ~=~ t \sum_{\ijc}~f_i^\dagger f_j \cF_{ij}
{}~-{J\over 4 }
\sum_{\ijkc}~(\delta_{ik}-f_k^\dagger f_i ) ~\cA_{ij}^\dagger \cA_{kj}~
(1-f_j^\dagger f_j)~~,
\eqno(1.7)
$$
where
$$
\eqalign{
\cA_{ij}^\dagger~=&
 a^\dagger_i b^\dagger_j - b^\dagger_i a^\dagger_j~, \cr
\cF^\dagger_{ij}~=&   a^\dagger_j a_i + b^\dagger_j
b_i ~.\cr
}
\eqno(1.8)
$$

It is now possible to generalize the t-J model to $S >1/2$ by replacing the
constraint (1.6) by
$$
a_i^\dagger a_i+ b_i^\dagger b_i+ f_i^\dagger f_i~=2S~~~~~\forall ~i.
\eqno(1.9)
$$
In the half-filled (undoped) case, $f^\dagger f~=0$, the Schwinger bosons
describe states of spin $S$, and their bilinear forms yield the matrix elements
of the standard spin operators
$$
\half (a_i^\dagger, b_i^\dagger) \sigma^\alpha
\pmatrix{a_i\cr
         b_i}~=\bS^\alpha_i~~~~,~~~\alpha=1,2,3.
\eqno(1.10)
$$
In the presence of a hole at site $i$ the spin size at that site gets reduced
by 1/2. We shall see that the large-S limit yields a non trivial classical
($S\to \infty$) theory provided we hold the scaled coupling constants
$\bJ,\bt$ fixed, where
$$
\bJ~=4JS^2 ~~~~, ~~~~~ \bt~=
2tS~.
\eqno(1.11)
$$
It is also possible to generalize the t-J model to large-N by introducing $N$
flavors of bosons per site [A2].
The large-N mean field
theory predicts spiral magnetic phases at finite doping
concentrations [J1]. Recently however, the fluctuation
determinant has  been found to be unstable (negative) in a {\it range
of momenta}. The offending fluctuations were identified as
local enhancements of the spiral distortion.
Clearly, the holes drive strong
perturbations of the spins on the lattice constant scale.
These are difficult to treat by direct application of
continuum and mean field approximations on the Hubbard and t-J models.
This is why we shall concentrate on the large-S expansion.

\beginsection 2. SPIN-HOLE COHERENT STATES

Spin coherent states have been fruitfully used by Haldane
to map the QHM onto the non linear sigma model [H1]. The construction of the
spin coherent states path integral allowed a simple derivation of the
topological Berry phases.  The large-S expansion of this path integral provides
a unified
semiclassical treatment of the ordered and disordered phases of the quantum
antiferromagnet. In the ordered phase, one obtains the usual spin wave
expansion about the N\'eel state, which yields a very good approximation for
the ground state staggered magnetization and uniform susceptibility even
for spin 1/2 [C1].

In this lecture we generalize this approach to represent the t-J
model by defining ``spin-hole coherent states''.
This will allow us to treat the short range interactions carefully, while
observing the local constraints. We derive a semiclassical
expansion of the ground state and low excitations in the presence of holes.
Although the expansion is formally controlled by the large spin size $S$, at
low doping we
may rely on the success of this approximation for the S=1/2 QHM.

Spin coherent states of spin $S$ are defined as
$$
|\OM_{S}(\theta,\phi)\rangle=  (2S!)^{-1/2}~(u a^{\dagger}
+ v
b^{\dagger})^{2S}|0,0\rangle~,
\eqno(2.1)
$$
where  $\OM$ is a unit vector, and
$$
u = \cos ( {\theta/ 2}) e^{i \phi /2 }~~~~~,~~~~v = \sin ( {\theta/ 2}) e^{-i
\phi /2 }~.
\eqno(2.2)
$$
$\theta,\phi$ are the lattitude and longitude angles on the unit sphere.

The spin-hole coherent states are defined as follows:
$$
|\OM,\xi \rangle_S~=~|\OM_{S}\rangle ~|0\rangle ~+~| \OM_{S-{1\over
2}}\rangle~\xi f^\dagger|0\rangle ~.
\eqno(2.3)
$$
where $\xi$ is an anticommuting Grassman variable.
The states (2.3) allow a resolution of the identity, while conserving the
constraint (1.9) for any given S-sector:
$$
{2S \over {4\pi}}\int d \phi ~ d \cos \theta ~d\xi^* d\xi \exp
\left[-\alpha_S~\xi^*
\xi \right] ~|\OM , \xi\rangle \langle \OM, \xi | ~=~ I ~.
\eqno(2.4)
$$
The factor $\alpha_S~=(2S+1)/(2S)$ is required for normalizing the
matrix elements to unity.
We shall be able to replace $\alpha$ by unity, by renormalizing  the chemical
potential $\mu$ in the grand cannonical ensemble.

By repeatedly inserting the resolution of the identity (2.4) in the Trotter
product expansion of the density matrix, we following the standard procedure
[H1] and construct the path integral
for the partition function:
$$
\eqalign{
Z~=&\int \cD \OM ~\cD\xi^* ~\cD\xi
\cr
&~~~~\exp \left[
\int_0^\beta
 d\tau~
\sum_i~\left(i (2S-\xi_i^* \xi_i ) \bA(\OM_i)\cdot{\dot \OM}_i +
\xi_i^* {\dot \xi}_i
\right)
{}~-H^{tJ} [\OM,\xi^* , \xi ]  \right]~,\cr
}
\eqno(2.5)
$$
where
$$
H^{tJ}~= {\langle \OM,\xi| \cH^{tJ}|\OM,\xi\rangle \over
\langle \OM,\xi|\OM,\xi\rangle}
\eqno(2.6)
$$
is a function of the variables $\theta_i,\phi_i,\xi_i$, and
$$
\bA \cdot{\dot\OM_i}~=(1-\cos\theta_i(\tau)){\dot \phi_i}(\tau)
\eqno(2.7)
$$
is the spin-kinetic term. The coordinate invariant notation uses the gauge
potential  $\bA({
\Omega})$ which describes the vector potential of a unit magnetic monopole at
the center of the sphere. The integral of the (2.7) over a closed loop
on the sphere yields the solid angle enclosed by that loop.

The Grassman ``time derivative'' is defined in its discrete form:
$$
{\dot\xi}= (\xi(\tau) - \xi(\tau-\epsilon))/\epsilon~,
\eqno(2.8)
$$
where $\epsilon$ is the infinitesimal timestep.

$H^{tJ}$ has quadratic and quartic terms in the Grassmann variables. It is
possible to integrate out the Grassmans exactly only for quadratic
hamiltonians. Therefore, we decouple the four-fermion
terms by a Hartree-Fock approximation. For our purposes
this approximation will be justified by the following arguments: (i)
hole-correlation
corrections to this Hartree-Fock decoupling are higher order in hole density
and (ii) the quartic terms vanish
in the ferromagnetically correlated regions, where the hole density will turn
out to be significant.

Thus we arrive at a model which is quadratic in the Grassman variables:
$$
\eqalign{
H^{tJ}~\approx&~~ {\bar H}^J~+\sum_{ij}\left({\bar
H}^f_{ij}-\delta_{ij}\mu\right)\xi^*_i\xi_j\cr
{\bar H}^J~=& - {{\bar J}\over 8}
\sum_\ijkc ~ (\delta_{ik}-e^{i\psi_{ik}}\rho_{ik}\rho_{jj})
\sqrt{ (1- \OM_{j} \cdot
\OM_{k}) (1- \OM_i \cdot \OM_{j})}~  \cr
{\bar H}_{ij}^f~=& {{\bar t}\over \sqrt{2}}
\sqrt{1+ \OM_i \cdot \OM_{j}}
{}~ e^{i\gamma_{ij}}\cr
&~+{{\bar J}\over 8}
\sum_k \sqrt{ (1- \OM_{i} \cdot
\OM_{k}) (1- \OM_k \cdot \OM_{j})}~\left( (1-\rho_{kk}) ~
        +  ( \delta_{ij} -1 )\right)~e^{i\psi_{ij} }\cr
}
\eqno(2.9)
$$
$\gamma_{ij}$ and $\psi_{ij} $ are the phases of $u^*_i u_j
+ v^*_i v_j$ and $(u^*_i v^*_k - v^*_i u^*_k)(u_j v_k - v_j u_k)$ respectively.
The hole density matrix $\rho_{ij}[\OM]~=\langle f^\dagger_i f_j \rangle$ is to
be determined self-consistently, by solving for the Fermion ground state in the
presence of
a spin configuration $\{\OM\}$.

The  hole hamiltonian ${\bar H}^f$ describes two distinct hopping processes:
inter-sublattice hopping
(``t-terms'') and intra-sublattice hopping (``J-terms'').
When the spins $\OM_i$ have short range
antiferromagnetic order the $\gamma$ phases in the t-terms fluctuate wildly,
while  $\psi_{ij}=\eta_i \bA^N\cdot(\bfx_{i}-\bfx_j)$ represents a slowly
varying
N\'eel gauge field $\bA^N(\bfx )$ which obeys
$$
\nabla \times \bA_i^N~=\half \bn_x \times \bn_y \cdot \bn~.
\eqno(2.10)
$$
$\bn \approx \eta_i \OM_i $ is the  staggered magnetization (N\'eel)
field, where
$\eta_i =+1, (-1)$ on sublattice
A (B) is the ``sublattice charge''. $\nabla \times \bA_i^N $ is the
``topological charge density''
whose integral for continuos fields on a compact surface,
is invariant [P1].

Weigmann, Wen, Shankar and Lee have studied Lagrangians which contain similar
intra-sublattice $\bA^N$-coupled hopping terms in the context of high $\Tc$
superconductivity [W1]. Their starting model differed however from the t-J
model in that it did not contain the inter-sublattice hopping terms (t-terms).
These terms {\it are not $\bA^N$-gauge invariant, and do not conserve the
sublattice charges}. The t-terms cannot be neglected,  especially in
the $\bt/\bJ >1$ regime. We shall see, however, that in the large-S limit
they are dynamically eliminated from the low energy and long wavelength
Lagrangian.

We begin by integrating out the fermions to obtain a purely
spin partition function
$$
  Z^s ~= \int \cD \OM   ~\exp  \left[ \int_0^\beta d\tau
\left(~i\sum_i (2S-\rho_i)\bA(\OM_i)\cdot {\dot\OM}_i ~- {\bar H}^J[\OM] -E^f
[\OM]\right)~\right]~,
\eqno(2.11)
$$
where
$$
E^f[\OM ]~=-\beta^{-1}{\rm Tr} ~\log\left[ 1 + T_\tau \exp\left[-\int_0^\beta
d\tau (H^f[\OM]-\mu)\right]  \right]
\eqno(2.12)
$$
is the fermion free energy in the presence  of a general spin history
$\OM(\tau)$. $T_\tau$ is the time ordering operator.
Here we concentrate on the zero temperature case $\beta=\infty$.
Eq. (2.11) is a useful starting point for the semiclassical
approximation. In the classical limit, $S\to\infty$,  the kinetic term is so
large, that the only important configurations in the path integral are
classical, i.e. the spins are frozen  and $\langle
{\dot \OM} \rangle =0$. One has therefore to  minimize ${\bar H}^J+E^f$ with
respect to $\OM$ for a given chemical potential or a specified
number of holes. The second step is include the semiclassical
fluctuations whose dynamics are controlled by the kinetic term.  We discuss the
single
hole and the many hole cases seperately.

\beginsection 3. RESULTS

In Ref. [A3],  a Lanzcos algorithm
on the Connection Machine was used to minimize the energy for $128\times
128$ spins. The following results were found:
\bigskip
\noindent
$\underline{The~single~hole}$:\hfil\break

In the regime $\bt/\bJ>0.87$,  we have found  that the
``polaron'' which is a {\it local} alignement of spins,  yields a lower energy
than any of the possible uniform states, including: the N\'eel state, spiral
states and canted states. This result helps to explain why the uniform spiral
states were found to be unstable against short wavelength distortions
in the mean field theory [A2].
The polaron variational parameters were chosen to describe a
ferromagnetic core, a transition region, and
a far field antiferromagnetic tail. The latter is completely determined by the
boundary condition $\delta\theta$  and the pure Heisenberg model (i.e. the
Laplace equation).

Our results are quite simple.  For $1< \bt/\bJ
< 4.1$ the single hole
energy is minimized by the five-site polaron (one flipped spin), depicted in
Fig. 1.

\midinsert            
\vskip 3 truein       
\centerline{{\bf Fig. 1:} {\sl The five site polaron. The hole density
$\rho_{ii}$ }}
\centerline{\sl is primarily concentrated on the sites of the unfilled arrows.}
\centerline{\sl The circular arrows represent an allowed tunneling path, where}
\centerline{\sl the polaron hops two lattice constants to the left.}
\endinsert
\bigskip
The hole density is
approximately $1/2$ and $1/8$ on the central and neigboring sites
respectively, with a small amount of leakage (due the J-terms in (2.9)) to
 sites further away. For $4.1 < \bt/\bJ  < 6.6$, the polaron has two flipped
spins
(diagonally across a plaquette), and at larger values the core radius increases
slowly as $~R_c~\sim (\bt/\bJ)^{1/4}$, and the energy goes
asymptotically as $\epsilon_h + 4\bt~\sim (\bJ\bt)^{1/2}$. The most
important fact is that the small polarons {\it do not distort the N\'eel
background}. In particular, the configurations centered on a
bond [A4] are considerably higher in energy.   We also find that
the polarons have no tails [S1], i.e. $\delta\theta=0$, throughout the regime
discussed above. This follows from competing contributions of order $~\pm
\bJ (\delta\theta)^2$ of ${\bar H}^J$ and $E^f$. Since in addition,  the
density $\rho$ is exponentially localized near the polaron sites,  we conclude
that
{\it the  classical interactions between polarons are short ranged}.

The polaron breaks the lattice translational symmetry. This symmetry is
restored by tunneling events, where two spins $i$ and $k$ simultaneusly flip
their orientation (see Fig. 1).  The
tunneling matrix element $\Gamma_{ik}$  (the polaron's hopping rate),
is non-perturbative in $S^{-1}$:
$$
\Gamma_{ik}=~\Gamma_0 \exp \left( -\sum_{i'} \int d{\tilde\phi_{i'}} {\bar
S}^z_{i'}\right) ~ \approx S^{1/2} \beta_{ik} \bt \exp( -S\alpha_{ik}) ~~,
\eqno(3.1)
$$
Eq. (3.1) can be computed as follows: The
azimuthal coordinates are analytically continued $\phi \to
{\tilde\phi}_i=\phi' + i \phi''$, while their cannonical momenta
$S^z_i=(2S-\rho_i)\cos\theta_i$ are  kept real. It can be readily verified
that $\sum_i S^z_i$, and ${\bar H}^J+E^f$ are conserved along the tunneling
path ${\bar S}^z_i({\tilde \phi})$ which minimizes the action. As a result of
these conservation laws, we obtain a selection rule:
{\it tunneling can only
take place between sites on the same sublattice!} This, in effect, amounts to
the elimination of the inter-sublattice ``t-terms''.

$\alpha_{ik}, \beta_{ik}$ are slowly varying dimensionless functions of
$\bt$ and $\bJ$.
For five site polarons, the number of spins involved in the tunneling path is
at least three. For
$S=\half$ we estimate the exponent to
be roughly unity, but a fuller treatment of the
multidimensional tunneling problem is
required for a quantitative determination of $\Gamma$ and the polaron's
effective mass.

The single
polaron in a perfect N\'eel background occupies a Bloch wave of dispersion
$\epsilon_\bk~=2\Gamma_{c}(\cos(2k_x)+cos(2k_y))+
2\Gamma_{b}(\cos(k_x+k_y)+cos(k_x-k_y))$, where $c,b$ denote the site of the
other flipped spin as labelled in Fig. 1. By energetic arguments,
$\Gamma_{b} \le \Gamma_{c}$.  A Berry's phase calculation of
the tunneling matrix elements for S=1/2, yields an overall positive sign for
$\Gamma_c, \Gamma_b > 0$. Thus the single polaron energy is minimized
at $\bk=(\pi/2,\pi/2)$. This result agrees with other studies
of the single hole spectral function in the t-J model [S2]. For
small deviations of the background  spins from antiferromagnetic
order the tunneling rate $\Gamma_{ik}$ is modulated by the
overlap of the background and the perfectly antiferromagnetic configurations.
This
overlap is just $\exp[i\eta_i\bA^N(\bfx_i-\bfx_k)]$. $\bA^N$ and $\eta_i$
are the aforementioned N\'eel gauge field and sublattice charge respectively.
{\it We notice that $\bA^N$ couples in a gauge invariant way to the polarons,
and that the sublattice charges are conserved in the hopping.}
\bigskip
\noindent
$\underline{Interactions}$:\hfil\break

The interactions between two polarons were computed in the regime
$\bt/\bJ=1-4$.  We define
$U_{ij}^p~=\half e_{ij}-2e_h$, where $e_{ij}$ is the relaxed energy of a
two-hole polaron with flipped spins at sites $i$ and $j$. $U_{ii}^p$
is repulsive, and of order
$0.6\bJ-2.6 \bJ$.  The intersite interactions, for
neigboring polarons at sites (a)-(d) (see Fig. 1.)  are plotted in Fig. 2.

\midinsert            
\vskip 2 truein       
\centerline{{\bf Fig. 2:} {\sl Classical interactions between polarons, in
units of $\bJ$. }}
\centerline{\sl Lines (a)-(d) represent the second polaron positions as
labelled in
Fig. 1.}
\centerline{\sl The solid line represents the relative condensation energy per
hole }
\centerline{\sl of thehole-rich phase (see text).  }

\endinsert
\bigskip

We find both attractive and repulsive interactions, and it is interesting to
note that
for $\bt/\bJ < 1.8$ there is a near neigbour attraction of
antiferromagnetically
correlated spins.  We also consider the
possibility of polaron condensation into hole-rich domains [I1].
The condensation energy per hole is determined by minimizing it with respect to
the
spin configuration, and the density. The spins in the hole-rich domains align
ferromagnetically, and the energy per hole is given by $e_{fm}~=
-4\bt + 4\sqrt{B\bJ\bt\pi}$. This results coincides with that of Emery,
Kivelson and
Lin [I1], except that their quantum correction factor B=0.584
is here set to
1/2. In Fig. 2., the condensation energy $\Delta e_c~=e_{fm}-e_h$ is plotted.
We find that it becomes negative at $\bt/\bJ=2.7$, {\it above which
phase-separation will occur for large S}.

Attractive interactions and negative {\it classical}
condensation energies may yield charge density waves or superconductivity in
the
ground state of the quantum model.
However, if realistic intersite Coulomb
repulsions are added to the t-J model, the short range interactions may change
sign.
In particular, phase separation may
be supressed, or pushed to higher values of $\bt/\bJ$.

The information given above allows us to write the effective
Lagrangian for a dilute system of small
polarons:
$$
\eqalign{
\cL^{s-p}~=&\sum_i \left[ i(2S-p^*_i p_i)
\bA(\OM_i)\cdot {\dot\OM}_i~+p_i^* {\dot
p}_i\right] ~+{\bJ\over 8}\sum_{\ijc}  \OM_i\cdot\OM_j ~
+\sum_i (e_h-\mu)~p^*_i p_i ~\cr
&~~+ \sum_\ijkc \Gamma_{ik} e^{\eta_i \bA^N \cdot
\bfx_{ik}}~ p^\dagger_i p_k ~+ \sum_{ij} U^p_{ij} ~p^*_i
p_i ~p^*_j p_j \cr}
\eqno(3.2)
$$
Eq. (3.2) is the most important result.
$\cL^{s-p}$ describes a two charge system of spinless
Fermions $p_i$ with short range interactions $U^p_{ij}$ coupled to Heisenberg
spins.
The
formation of polarons can be viewed
as a strong short-wavelength dressing of the
original f-holes by the spins.
As a consequence, the uncomfortable $t$-terms have
been conveniently eliminated, and the effects of holes on the spin background
is
short ranged. This Lagrangian describes the low lying excitations of the
so-called t'-J Hamiltonian
$$
\cH^{t'J}~={\bJ\over 2}\sum_{\ijc}  \bS_i\cdot\bS_j+ 4\sum_\ijkc \Gamma_{ik}
p^\dagger_i p_k \cA_{ij}^\dagger \cA_{jk}
\eqno(3.3)
$$
A major advantage of the  t'-J  model over the t-J model is that
in the small concentration limit $\delta << 1$, (3.2) is amenable to the
continuum approximation. Following Haldane [H1] the spin interactions can
be relaced by the (2+1) dimensional non linear sigma model, with
$\delta$ dependent renormalized stiffness constant and spin wave velocity.  The
precise evaluation of the sigma model parameters for finite $\delta$ is beyond
the
scope of this paper, but we expect that above some critical density
$\delta>\delta_c$ the ground state is disordered [C2], i.e. a ``spin liquid''.
In the massive ``spin liquid'' phase, Eq. (3.2) reduces to Wiegmann, Wen ,
Shankar and Lee's model [W1]:
$$
\eqalign{
\cL^{WWSL}~=&  \sum_{\eta=\pm  1}
\left[ p_\eta^\dagger ( \partial_\tau +
i \eta A^N_0 ) p_{\eta}~+
{1\over 2m} p^\dagger_{\eta} |\nabla + i\eta \bA^N_0 |^2 p_{\eta}\right] +
{1\over 4 \kappa} (F_{\mu\nu})^2\cr
&~~~~~~+\cO (p^\dagger p p^\dagger p) \ldots\cr
}
\eqno(3.4)
$$
where $m$ is the effective mass at $\bk=(\pi/2,\pi/2)$, and the
``electromagnetic'' N\'eel-fields are $F_{\nu\mu}= (\partial_\mu A^N_\nu -
\partial_\nu A^N_\mu)$. $\kappa$ is the inverse spin correlation length, which
is
also the coupling constant of the gauge field.  Previous analyses [W1] have
concluded that the ground state of (3.4) is most likely  an  RVB-type
superconductor.
Lee argued that the pairing is caused by two effects: (i) attraction
between the opposite charges induced by the N\'eel gauge field, and (ii)
suppression
of coherent single particle propagation due to fluctuating Bohm-Aharonov
phases,
while the pairs $\langle p^\dagger_{+}p^\dagger_{-}\rangle$ propagate as free
bosons. Both (i) and (ii) are only valid in the magnetically disordered phase,
a
pleasing feature which agrees with  the phase diagrams of the copper oxide
superconductors.

Aside from the mechanism of superconductivity, the small
polaron theory could be checked numerically on finite lattice
Monte-Carlo simulations, and experimentally in the copper-oxides and other
doped
antiferromagnets.  For example: the polaron size can be estimated by
NMR techniques [M1], and its internal excitations could be probed by optical
absorption.  In the frozen moments regime, one expects the polarons to  exhibit
conductivity typical of weakly localized  semiconductors [C3].

\beginsection \underbar{\it Acknowledgements}

The author is grateful to B. Larson with whom part of this work was done, and
to S. Kivelson for many fruitful discussions. This work is supported in part by
a grant from the US-Israel Binational Science Foundation.
\vfill\eject

\beginsection REFERENCES

\hyphenpenalty=10000
\def\ref#1\par{\parshape=2.0in 15.2 truecm 1.0 truecm 14.2 truecm
{#1}\par}
\parskip=0pt
\parindent=0pt

\ref
[A1] Auerbach,~A. and Arovas,~D.P.,  Phys. Rev. Lett. {\bf 61}, 617 (1988);
Jour. Appl. Phys. {\bf 67}, 5734  (1990).

\ref
[A2] Auerbach,~A. and Larson,~B.E.,  Phys. Rev. B{\bf 43}, 7800 (1991).

\ref
[A3]   Auerbach,~A. and Larson,~B.E.,  Phys. Rev. Lett. {\bf 66}, 2262  (1991).

\ref
[A4] Aharony,~A.,Birgeneau,~R.J., Coniglio,~A., Kastner,~M.A.  and
Stanley,~H.E.,
\prl {\bf 60}, 1330 (1988).

\ref
[C1] Chakravarty~S., Proc. of 1989 Symposium on High Tc
Superconductivity, Ed. Bedell,~K.S.  {\it et. al.}, (Addison-Wesley, 1990) and
references
therein.

\ref
[C2] Spin wave theory about a single polaron yields
a finite $\cO(\delta)$ correction to the ground state magnetization. This
suggests
that $\delta_c \ne 0$.

\ref
[C3] Chen~C.Y. {\it et. al}, \prl {\bf 63}, 2307 (1989).

\ref
[H1] Haldane,~F.D.M. , ``Two Dimensional Strongly Correlated Electron
Systems'',
Eds. Gan,~Z.Z.  and Su,~Z.B. , (Gordon and Breach, 1988), pp. 249-261;\prl {\bf
61}, 1029 (1988).

\ref
[I1] Ioffe,~L.B.  and Larkin,~A.I. , \prb {\bf 37}, 5730
(1988);
Emery,~V.J., Kivelson,~S.A.  and Lin,~H.Q. ,  Phys. Rev. Lett. {\bf  64}, 475,
(1990); Marder,~M., Papanicolau,~N.  and Psaltakis,~G.C. ,  Phys. Rev. B{\bf
41}, 6920  (1990).

\ref
[J1] Jayaprakash,~C., Krishnamurthy,~H.R.,  and Sarker,~S. , \prb {\bf  40},
2610, (1989);
Kane,~C.L., Lee,~P.A. , Ng,~T.K., Chakraborty,~B.  and Read,~N. ,  \prb{\bf
41}, 2653
(1990).

\ref
[M1] Mendels,~P., {\it et. al.} Physica {\bf C 171}, 429 (1990).

\ref
[P1] Polyakov,~A.~M. , ``Gauge Field And Strings'', (Harwood, 1987), P. 140.

\ref
[S1] Schraiman,~B.I.  and Siggia,~E.D. , \prl {\bf 62}, 156 (1989).

[\ref
W1] Wiegmann,~P.~B., \prl {\bf 60}, 821  (1988); Wen,~X-G., \prb {\bf 39}, 7223
(1989); Shankar,~R., \prl {\bf 63}, 203  (1989); Lee,~P.A., \prl {\bf 63}, 690,
(1989).

\bye